# Detailed Study of Stars and Gas in a $z = 8.3$ Massive Merger with Extreme Dust Conditions

Anishya Harshan[1], Roberta Tripodi[1], Nicholas S. Martis[1], Gregor Rihtaršič[1], Maruša Bradač[1,2], Yoshihisa Asada[3,4], Gabe Brammer[5,6], Guillaume Desprez[3], Vince Estrada-Carpenter[3], Jasleen Matharu[5,6], Vladan Markov[1], Adam Muzzin[7], Lamiya Mowla[8,9], Gaël Noirot[3], Ghassan T. E. Sarrouh[7], Marcin Sawicki[3], Victoria Strait[5,6], and Chris Willott[10]

[1] University of Ljubljana, Department of Mathematics and Physics, Jadranska ulica 19, SI-1000 Ljubljana, Slovenia
[2] Department of Physics and Astronomy, University of California Davis, 1 Shields Avenue, Davis, CA 95616, USA
[3] Department of Astronomy and Physics and Institute for Computational Astrophysics, Saint Mary's University, 923 Robie Street, Halifax, Nova Scotia B3H 3C3, Canada
[4] Department of Astronomy, Kyoto University, Sakyo-ku, Kyoto 606-8502, Japan
[5] Cosmic Dawn Center (DAWN), Denmark
[6] Niels Bohr Institute, University of Copenhagen, Jagtvej 128, DK-2200 Copenhagen N, Denmark
[7] Department of Physics and Astronomy, York University, 4700 Keele St. Toronto, Ontario, M3J 1P3, Canada
[8] Whitin Observatory, Department of Physics and Astronomy, Wellesley College, 106 Central Street, Wellesley, MA 02481, USA
[9] Dunlap Institute for Astronomy and Astrophysics, 50 St. George Street, Toronto, Ontario M5S 3H4, Canada
[10] NRC Herzberg, 5071 West Saanich Rd, Victoria, BC V9E 2E7, Canada
Received 2024 August 21; revised 2024 November 23; accepted 2024 November 24; published 2024 December 12

## Abstract

We present galaxy MACS0416-Y1 at $z_{\rm spec} = 8.312$ as observed by the CAnadian NIRISS Unbiased Cluster Survey. MACS0416-Y1 has been shown to have extreme dust properties; thus, we study the physical properties and star formation histories of its resolved components. Overall, we find that MACS0416-Y1 is undergoing a star formation burst in three resolved clumps. The central clump is less massive compared to the other clumps and possibly formed in the merging process of the two larger clumps. Although the star formation history indicates an ongoing star formation burst, this gas-rich galaxy shows comparable star formation efficiency to cosmic noon galaxies. Using NIRSpec prism spectroscopy, we measure metallicity, $12 + \log(\mathrm{O/H}) = 7.76 \pm 0.03$, ionization parameter, $\log U = -2.48 \pm 0.03$, and electron temperature $T_e = 18000 \pm 4000$ K. The emission line ratios of the galaxy indicate an evolved interstellar medium similar to $z \sim 2$ star-forming galaxies. Further, we find possible presence of ionization from an active galactic nucleus (AGN) using emission line diagnostics; however, we do not detect a broad-line component in the H$\beta$ emission line. As this gas-rich galaxy is undergoing a major merger, we hypothesize that the high dust temperature in MACS0416-Y1 is caused by the star formation burst or a possible narrow-line AGN.

*Unified Astronomy Thesaurus concepts:* High-redshift galaxies (734); Strong gravitational lensing (1643); Reionization (1383); Active galaxies (17)

## 1. Introduction

How the first galaxies formed and evolved in the early Universe is one of the fundamental questions of extragalactic astronomy. As deep extragalactic surveys with JWST reveal the interstellar medium (ISM) and star formation properties of young galaxies, how massive galaxies formed at very early epochs has become a central question. Many existing studies explore the properties and physical scenarios to explain the formation of massive galaxies in in the first billion years after the Big Bang (e.g., E. Curtis-Lake et al. 2023; S. Fujimoto et al. 2023; B. E. Robertson et al. 2023; M. Xiao et al. 2024; C. M. Casey et al. 2024; G. Desprez et al. 2024). While these studies primarily rely on studying the integrated physical properties of galaxies, a resolved view of the ISM and star formation would provide us with insight into the formation and evolution of such massive galaxies in early epochs (e.g., Y. Asada et al. 2023; G. C. Jones et al. 2024a; E. Parlanti et al. 2024).

MACS0416-Y1 was first identified using multiband Hubble Space Telescope (HST)/WFC3 imaging by L. Infante et al. (2015) and N. Laporte et al. (2015) as a Lyman break galaxy candidate in the Hubble Frontier Field (HFF) lensing cluster MACS0416 (N. Laporte et al. 2015). It has been an object of further exploration using the Spitzer Infrared Array Camera, the Atacama Large Millimeter/sub-millimeter Array (ALMA) 850 μm, and the Jansky Very Large Array (JVLA; N. Laporte et al. 2015; G. B. Brammer et al. 2016; Y. Tamura et al. 2019; G. C. Jones et al. 2024b). Y. Tamura et al. (2019) and T. J. L. C. Bakx et al. (2020) confirm the redshift of MACS0416-Y1 at $z = 8.312$ using [O III] 88 μm and [C II] 158 μm emission lines. Y. Tamura et al. (2019) estimated its lensing-corrected stellar mass to be $\log M^*/M_\odot = 9.5$ and molecular gas mass to be $10^{10} M_\odot$ and described MACS0416-Y1 as a starburst galaxy. T. J. L. C. Bakx et al. (2020), using a low-resolution [C II] measurement, find a rotational gradient and three star formation clumps, also seen in rest-UV imaging in Y. Tamura et al. (2023). G. C. Jones et al. (2024b), using JVLA observations of upper limits of CO(2–1) and rest-frame radio continuum emission, find an extremely high dust temperature of $>90$ K, indicating extreme ISM conditions in MACS0416-Y1.

Recently, MACS0416-Y1 has been observed as part of the CAnadian NIRISS Unbiased Cluster Survey (CANUCS; C. J. Willott et al. 2022). G. Desprez et al. (2024) confirm the spectroscopic redshift of MACS0416-Y1 using a NIRSpec prism spectrum. In this Letter, using the CANUCS NIRCam + NIRSpec rest-frame UV + optical observations along with existing observations from ALMA, we present the resolved ISM and stellar properties of MACS0416-Y1.

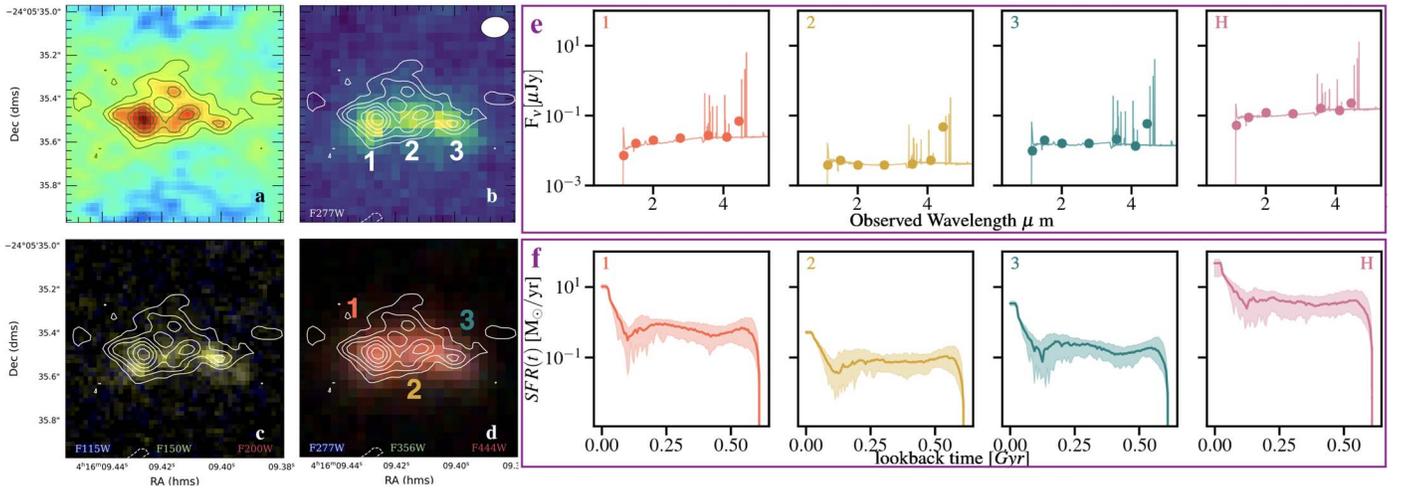

**Figure 1.** MACS0416-Y1 in multiwavelength imaging. (a) The 353 GHz dust continuum emission map obtained from B7 ALMA observations of MACS0416-Y1. Contours are drawn at 2, 3, 4, 5, 6, 7$\sigma$ with $\sigma = 5$ $\mu$Jy beam$^{-1}$. The same contours are overlaid on NIRCam imaging. (b) Image in NIRCam F277W, which shows the three stellar components of the galaxy along with the ALMA beam. (c) Red, green, and blue (RGB) image in UV using NIRCam F115W, F150W, and F200W. The UV emission is offset from the dust continuum showing dust obscured star formation in the galaxy. (d) RGB image in optical using NIRCam F277W, F356W, and F444W. The three clumps are marked. (e) The observed PSF-homogenized photometry of resolved components of the galaxy along with the best-fit model SED from DENSE BASIS. (f) The nonparametric star formation histories of the three resolved clumps and the galaxy halo inferred using DENSE BASIS. All components show older stellar populations and a recent burst, with the majority of the stars being formed in the galaxy halo.

We assume a $\Lambda$CDM cosmology with $H_0 = 70$ km s$^{-1}$ Mpc$^{-1}$, $\Omega_m = 0.3$, and $\Omega_\Lambda = 0.7$. All magnitudes are AB magnitudes.

## 2. Data and Methodology

We use the data acquired by the CANUCS NIRISS Guaranteed Time Observations Program ID 1208 (C. J. Willott et al. 2022), which covers five different strong lensing cluster fields including MACS J0416.1-2403 (hereafter MACS0416, $z = 0.395$). Each of the five cluster fields is accompanied by a NIRCam flanking field and a NIRISS flanking field. All five clusters are followed up with low-resolution NIRSpec prism spectroscopy with microshutter assembly (MSA) mode.

### 2.1. Imaging and Resolved Photometry

MACS0416 was observed with NIRCam filters F090W, F115W, F150W, F200W, F277W, F356W, F410M, and F444W with exposure times of 6.4 ks each. We also utilized archival data of HST imaging from the HFF program (J. M. Lotz et al. 2017). The CANUCS image reduction and photometry procedure is described in detail in Y. Asada et al. (2024). In short, we use a modified version of the Detector1-Pipeline (calwebb_detector1) stage of the official STScI pipeline and jwst_0916.pmap JWST Operational Pipeline (CRDS_CTX) to reduce the NIRCam data. We perform astrometric alignment of the different exposures of JWST/NIRCam to HST Advanced Camera for Surveys images, sky subtraction, and drizzling to a common pixel scale of 0″.04 using version 1.6.0 of the grism redshift and line analysis software for space-based slitless spectroscopy (G. Brammer & J. Matharu 2021, Grizli). The source detection and photometry are done with the Photutils package (L. Bradley et al. 2022) on the $\chi_{mean}$ detection image created using all available NIRCam images.

We performed photometry on MACS0416-Y1 in the six JWST bands (F150W, F200W, F277W, F356W, F410M, and F444W) where the galaxy is detected, using the technique described in M. Brada et al. (2024) and L. Mowla et al. (2024). The galaxy is clearly resolved into three separate clumps and a smooth component (halo; Figures 1(b), (d)). We extracted the point-spread functions (PSFs) of each filter empirically following the method described in G. T. E. Sarrouh et al. (2024).

We created 7″ × 7″ images centered on the galaxy in all six filters and visually determined the priors for the centers of the three clumps and the halo. We fit the central coordinates of the four components using three point sources (for the clumps) and a Sérsic profile (for the halo) in the F277W image with GALFIT (C. Y. Peng et al. 2010). These central coordinates were then used as fixed priors to repeat the process in all six filters and extract the photometry of the four components. To derive the uncertainty in our flux estimation, we injected the model of MACS0416-Y1 into 100 random positions in our 7″ × 7″ images and refitted them with the same GALFIT settings. We found no significant systematic offset between the fitted and injected fluxes of any component, demonstrating the robustness of our photometric technique. Further, we fit Sérsic profiles for resolved clumps 1 and 3 with fixed positions and similar parameters to get the approximate radii of the clumps and the integrated galaxy. The magnification-corrected ($\mu = 1.60^{+0.01}_{-0.02}$; refer to Section 2.4) radii of the four components and the galaxy are provided in Table 1.

### 2.2. NIRSpec Spectroscopy

CANUCS observations also consist of NIRSpec prism spectra ($R = 100$). MACS0416-Y1 has been observed with two three-shutter slits (Figure 2(a)). NIRSpec data have been reduced using the JWST pipeline for stage 1 corrections and then the msaexp (G. Brammer 2022) to create wavelength-

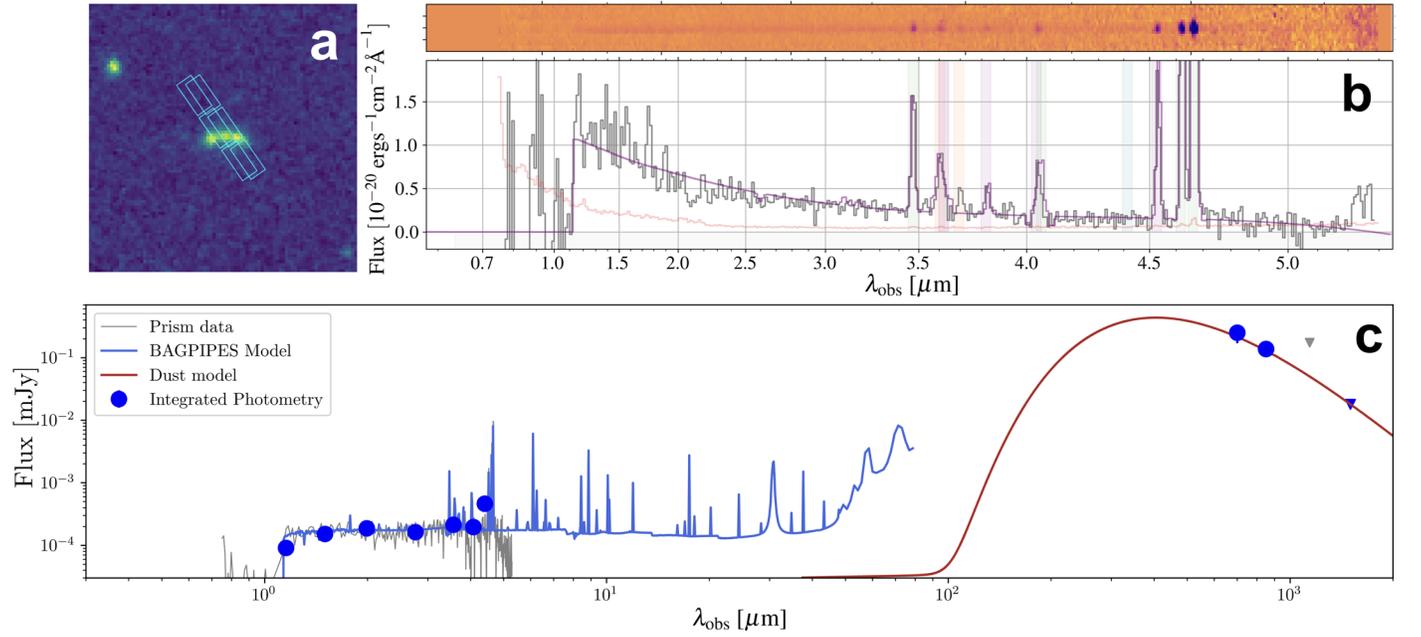

**Figure 2.** The 2D and 1D spectrum from the NIRSpec MSA in two slits overlaid on the MACS0416-Y1 NIRCam F277W image (panel (a)). Panel (b) shows the interpolated model of the continuum used and emission lines fits in purple, the 1D spectrum is shown in black, and the error is shown in red. Panel (c) shows the spectrophotometric fit to integrated NIRCam photometry (blue circles) and NIRSpec prism spectrum (gray line) from Bagpipes (blue line) and the dust emission model fit to ALMA photometry (red).

**Table 1**
Physical Properties of MACS0416-Y1 from DENSE BASIS, Bagpipes, and GALFIT

| Object | $R_e$ (kpc) | log $M_*/M_\odot$ | log SFR ($M_\odot$ yr$^{-1}$) | 12+log(O/H) | $A_V$ | T50 (Gyr) |
|---|---|---|---|---|---|---|
| | | | DENSE BASIS | | | |
| Integrated photometry | 1.19 ± 0.05 | 9.0 ± 0.07 | 1.5 ± 0.04 | 7.86 ± 0.07 | 0.39 ± 0.04 | 0.18 ± 0.1 |
| Clump 1 | 0.31 ± 0.05 | 8.6 ± 0.3 | 0.2 ± 0.5 | 7.89 ± 0.50 | 0.52 ± 0.3 | 0.31 ± 0.1 |
| Clump 2 | < 0.2 | 7.6 ± 0.3 | − 0.4 ± 0.3 | 7.68 ± 0.42 | 0.19 ± 0.13 | 0.27 ± 0.1 |
| Clump 3 | 0.31 ± 0.08 | 8.2 ± 0.2 | 0.2 ± 0.3 | 7.94 ± 0.45 | 0.21 ± 0.14 | 0.26 ± 0.1 |
| Halo | 1.48 ± 0.54 | 9.4 ± 0.1 | 0.8 ± 0.5 | 7.78 ± 0.51 | 0.36 ± 0.2 | 0.27 ± 0.1 |
| Integrated Spectrophotometry | | | Bagpipes | | | |
| dblplaw SFH | | 8.88 ± 0.08 | 1.84 ± 0.08 | 8.11 ± 0.26 | 0.59 ± 009 | … |
| K. G. Iyer et al. (2019) SFH | | 8.90 ± 0.07 | 1.92 ± 0.06 | 8.04 ± 0.11 | 0.58 ± 006 | 0.29 ± 0.05 |

**Table 2**
ISM Properties, Emission Line Ratios, and and Emission Fluxes

| Galaxy Property | Integrated MACS0416-Y1 | Galaxy Property | Integrated MACS0416-Y1 |
|---|---|---|---|
| Flux [O III] 5007 | 251.6 ± 1.6 | RO3 | 1.70 ± 0.06 |
| Flux [O III] 4969 | 94.2 ± 1.3 | O32 | 0.63 ± 0.01 |
| Flux H$\beta$ | 36.2 ± 1.1 | Ne3O2 | −0.60 ± 0.05 |
| Flux [O III] 4363 | 5.6 ± 0.6 | O3Hg | −0.45 ± 0.08 |
| Flux H$\gamma$ | 15.8 ± 1.0 | O3O3 | 1.56 ± 0.05 |
| Flux H$\delta$ | 8.0 ± 0.7 | H$\gamma$/H$\beta$ | 1.92 ± 0.10 |
| Flux [Ne III]3868 | 9.9 ± 1.0 | Z (12+log(O/H)) | 7.76 ± 0.03 |
| Flux [O II] 3727+3729 | 37.8 ± 1.0 | $T_e$ | 17634 ± 3900 |
| R3 | 0.83 ± 0.01 | log $U$(O32) | −2.48 ± 0.30 |
| R2 | 0.19 ± 0.02 | log $U$(Ne3O2) | −3.10 ± 0.30 |
| R23 | 0.91 ± 0.03 | $E(B−V)$ | 0.349 |

**Note.** Fluxes in $10^{-20}$ ergs cm$^{-2}$ s$^{-1}$ $T_e$ in K.

calibrated, background-subtracted 2D spectra. We extract the 2D spectrum for the combined slits. We fit a Gaussian along the spatial axis of the 2D spectrum and extract the 1D spectrum. We fit each 1D spectrum to a spline model using msaexp to create the continuum model and extract fluxes, velocity offsets, and widths of emission lines from the continuum-subtracted spectrum (Figure 2(b)). We report the fluxes and emission line ratios in Table 2.

### 2.3. ALMA Data Reduction and Analysis

We use all the data presented in Table 1 of Y. Tamura et al. (2023) to reproduce the contours of the continuum emission in ALMA Band 7 (B7) at 353 GHz (Figure 1). Additionally, we analyze the previously unpublished data set 2019.1.00343.S designed to detect the continuum emission of MACS0416 in ALMA Band 8 (B8) at 465 GHz. The visibility calibration of the observations was performed by the ALMA science archive. The imaging was performed through the Common Astronomy Software Applications (CASA Team et al. 2022), version 6.6.0-20. We applied the task tclean using natural weighting and a $3\sigma$ cleaning threshold. We image the continuum in B7 and B8 using the multifrequency synthesis mode in all line-free channels, selected by inspecting all the spectral windows.

For B8 observation, the resulting continuum image has an rms of 68 $\mu$Jy beam$^{-1}$ and a resolution of $(0.84 \times 0.63)$ arcsec$^2$ with a position angle of 72°. We analyze the new continuum emission in B8 at 465 GHz, and by performing a 2D Gaussian fit, we derive a flux density of $254 \pm 81$ $\mu$Jy and a peak flux of $199 \pm 34$ $\mu$Jy ($6\sigma$ detection). The source is barely resolved with a size of $(0.52 \pm 0.39) \times (0.21 \pm 0.31)$ arcsec$^2$.

### 2.4. Lens Modeling

For the magnification estimate, we use the cluster-scale lens model presented in G. Rihtaršič et al. (2024), constructed with public lens modeling software Lenstool (E. Jullo et al. 2007). The updated catalog of model constraints includes 303 multiple images with spectroscopic redshifts—the largest such data set to date. This includes systems with previously known MUSE redshifts and systems for which we obtained spectroscopic redshift for the first time using NIRISS and NIRSpec spectroscopy. The magnification of MACS0416-Y1 derived from the lens model is $\mu = 1.60^{+0.01}_{-0.02}$. The tangential and radial magnifications are small ($1.46 \pm 0.01$ and $1.08 \pm 0.01$, respectively) and do not significantly change the shape of the image.

### 2.5. Cold Dust Spectral Energy Distribution Fitting

The availability of additional ALMA B8 observations allows us to better constrain the dust properties of MACS0416-Y1 compared to existing studies. We perform a fit of the cold dust spectral energy distribution (SED) using the continuum emission detected at 465 and 353 GHz from Y. Tamura et al. (2019) and the stringent $3\sigma$ upper limit at 200 GHz from T. J. L. C. Bakx et al. (2020). We note that, by including the stringent upper limit in B3, we derive dust masses about 1 order of magnitude smaller than that of Y. Tamura et al. 2019 (adopting $\beta = 1.5$ and $T_{\rm dust} = 50$ K). We do not use the upper limit at 260 GHz found in Y. Tamura et al. (2019) since it is shallower compared to the other observations (see the gray triangle in Figure 2(c)), and therefore, it cannot put further constraints on the dust properties. MACS0416-Y1 also has shallow observations at lower frequencies (see also G. C. Jones et al. 2024b). Therefore, we assume a small radio contribution and model it as done in G. C. Jones et al. (2024b), adopting the nonthermal emission model of H. S. B. Algera et al. (2021).

We model the dust continuum emission with a modified black body (MBB) function. Details about the fitting functions and procedure can be found in the Appendix B. Given that the lack of data in the far-infrared prevents us from estimating the dust temperature ($T_{\rm dust}$), we create models using two free parameters, the dust mass ($M_{\rm dust}$) and the dust emissivity index ($\beta$), at different fixed $T_{\rm dust}$. Furthermore, we force the model to reach the flux of the 200 GHz upper limit. We find that fixing $T_{\rm dust}$ at values <60 K results in nonphysical values of $\beta$ ($\beta > 2.6$) ($\beta = 1.6$ typically found in star-forming and QSO's host galaxies; see e.g., J. Witstok et al. 2023; R. Tripodi et al. 2024). Figure 2(c) shows the best-fitting model with $T_{\rm dust} = 60$ K, resulting in $\beta = 2.6$ and $\mu M_{\rm dust} = 2.8 \times 10^5 M_\odot$, and $\mu$SFR = 78 $M_\odot$ yr$^{-1}$ (adopting a Chabrier initial mass function, IMF; G. Chabrier 2003). Considering the strong emission from [C II] and [O III] lines (Y. Tamura et al. 2019; T. J. L. C. Bakx et al. 2020), the implied star formation rate (SFR) is of the order of $\sim 60-100 M_\odot$ yr$^{-1}$. The best-fitting model with $T_{\rm dust} = 80$ K, having $\beta = 2.3$ and $\mu M_{\rm dust} = 3.4 \times 10^5 M_\odot$, yields $\mu$SFR = 167 $M_\odot$ yr$^{-1}$, which is a factor of 1.7 higher than the upper bound for SFR from emission lines. Therefore, the dust temperature of MACS0416-Y1 may be in the range 60–80 K.

The new measured dust temperature for MACS0416-Y1 at 60–80 K is higher compared to the dust temperature of star-forming galaxies at $z \sim 2$, which are reported to have $T_{\rm dust} \sim 30$ K (R. M. Mérida 2024) and $T_{\rm dust} \sim 30-45$ K at $z \gtrsim 5$ (A. L. Faisst et al. 2020; Y. Khusanova et al. 2021; L. Sommovigo et al. 2021; J. Witstok et al. 2023; H. Algera et al. 2024). However, our results have lower dust temperature compared to the findings of M. P. Viero et al. (2022), who find $T_{\rm dust} \sim 100 \pm 2$ K at $z \sim 7$. The sample of star-forming galaxies reported in M. P. Viero et al. (2022) at $z > 6$ is based on shallow imaging and extremely high stellar mass, indicating a large fraction of low-redshift interlopers. Indeed, using the same data set with robust photometric cuts, G. T. Jones & E. R. Stanway (2023) found the evolution of $T_{\rm dust}$ between $0 < z < 5$ to be much shallower than M. P. Viero et al. (2022). The dust temperature measured for MACS0416-Y1 is in agreement with the extrapolation of the relation reported in G. T. Jones & E. R. Stanway (2023) and with the wide range of dust temperatures previously reported in T. J. L. C. Bakx et al. (2020), G. C. Jones et al. (2024b), and L. Sommovigo et al. (2021).

### 2.6. Optical–Near-infrared SED Fitting

We perform two sets of SED fitting in order to form a complete picture of this complex system. First, we use DENSE BASIS (K. G. Iyer et al. 2019) to infer the stellar population properties of each of the resolved components by fitting the resolved photometry. V. Markov et al. (2023, 2024) find that the dust attenuation curves of galaxies at $z \gtrsim 8$ are Calzetti like; thus, we use DENSE BASIS with the D. Calzetti et al. (2000) dust attenuation and G. Chabrier (2003) IMF. We also perform a spectrophotometric fit to the integrated NIRCam photometry and NIRSpec spectroscopy using Bagpipes (A. C. Carnall et al. 2018, 2019). The spectrum is scaled to the photometry to account for slit losses. The Bagpipes fit is fixed to the spectroscopic redshift of 8.31 and assumes a double power-law star formation history (SFH), D. Calzetti et al. (2000) dust attenuation, and G. Chabrier (2003) IMF. We also performed a fit with the nonparametric SFH from K. G. Iyer et al. (2019) to enable a more direct comparison with DENSE BASIS and found similar results to the double power-law SFH. Figure 2 shows the Bagpipes spectrophotometric fit in blue, thermal dust emission model in red, and observed photometry in blue.

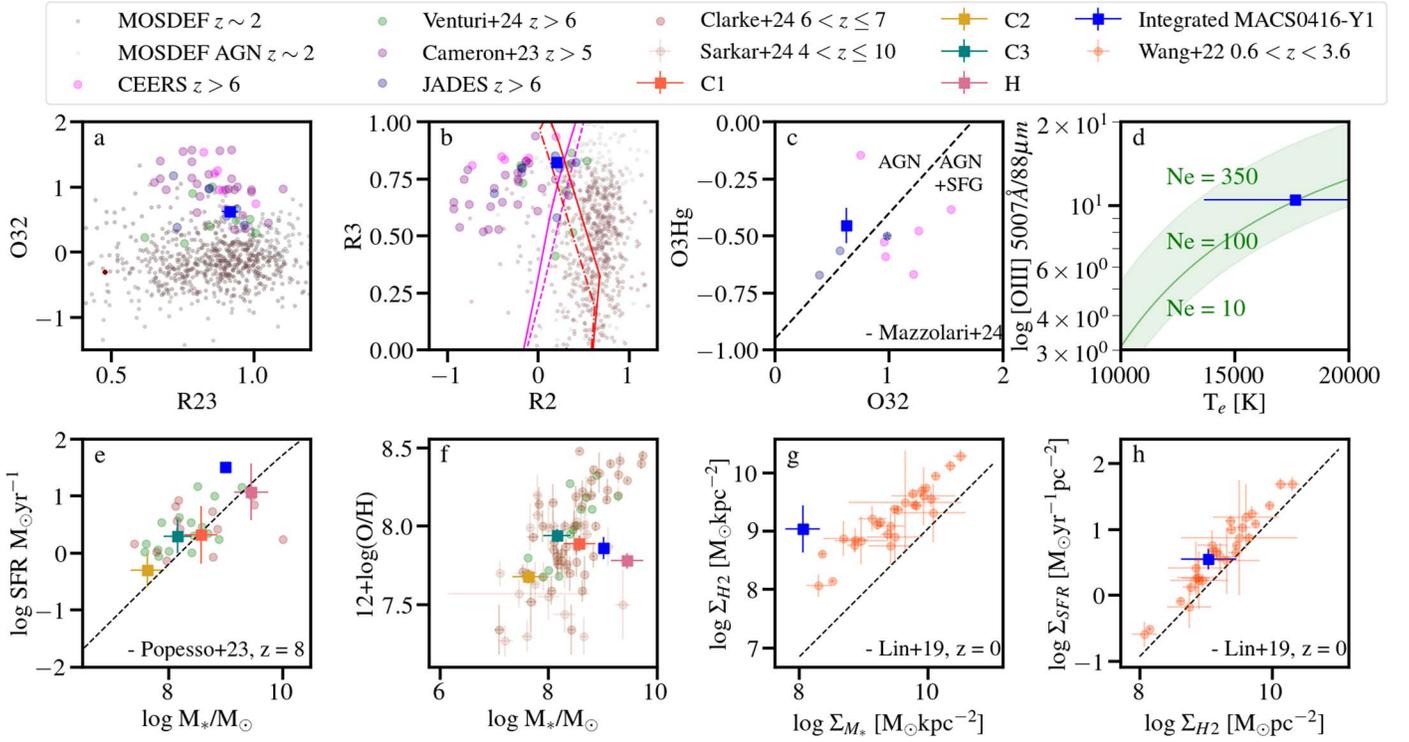

**Figure 3.** (a) O32-R23 diagram using dust-corrected emission line ratios. (b) R3-R2 diagram, overlaid with shock (M. G. Allen et al. 2008) and AGN (B. A. Groves et al. 2004) models from MAPPINGS in pink and red, respectively (described in Section 3.3). (c) O3Hg-O32 AGN diagnostic from G. Mazzolari (2024) indicating presence of AGN. (d) Electron density diagnostic from PyNeb indicates that the electron density of MACS0416-Y1 is between 10 and 350 cm$^{-3}$. (e) SFR vs. stellar mass of the four resolved components of the galaxy and the integrated galaxy with SFR main sequence from P. Popesso et al. (2023) at $z = 8$. (f) Stellar mass–metallicity for the four resolved components of the galaxy and the integrated galaxy with metallicity taken from DENSE BASIS SED fitting. (g) Molecular gas surface density (log $\Sigma_{H_2}$)–stellar mass surface density (log $\Sigma_{M_*}$) of the integrated galaxy show that MACS0416-Y1 is more gas rich compared to cosmic noon and local galaxies. (h) SFR surface density (log $\Sigma_{SFR}$)–molecular gas surface density (log $\Sigma_{H_2}$) plot of the integrated galaxy show that MACS0416-Y1 has similar SFE to cosmic noon galaxies. In all panels, error bars where not visible are small.

The fitting parameters and fitted results for Bagpipes and DENSE BASIS are summarized in Table 1.

## 3. Results

### 3.1. Resolved SFRs and Histories

As described in Section 2.6, we perform SED modeling for the three resolved clumps and the underlying galaxy halo using DENSE BASIS and construct the nonparametric star formation histories shown in Figure 1.

The star formation histories of the resolved components show two episodes of star formation prominent in the galaxy. All components have formed 50% of their stars in the last 300 Myr. All components are also undergoing a simultaneous burst of star formation in the last 150 Myr (bottom-right panels in Figure 1). The simultaneous burst of star formation in the galaxy components indicate that MACS0416-Y1 is a late-stage merger.

We further measure the UV slope ($\beta_{UV}$) of each component from photometry as $\beta_{UV} = -1.26 \pm 0.2$, $-3.06 \pm 0.3$, $-2.66 \pm 0.1$, and $-1.02 \pm 0.1$ for clumps 1, 2, 3, and H, respectively. Similarly, we measure the $\beta_{UV} = -1.26 \pm 0.1$ for the integrated photometry comparable to $\beta_{UV} = -1.19 \pm 0.2$ from spectroscopy. The underlying galaxy halo H's and clump 1's shallow UV slope indicates an older stellar population and high dust attenuation. These results are comparable to Y. Tamura et al. (2019), who found that the MACS0416-Y1 has two distinct stellar populations and is currently experiencing a star formation burst. On the other hand, Z. Ma et al. (2024)

find that the three components and the integrated MACS0416-Y1 are going through extreme first starburst with average stellar ages of $< 10$ Myr and very high SFR. These star formation histories require very fast metal enrichment to explain the metallicity and high dust mass.

In Figure 3(e), we show that the resolved components of the galaxy and the integrated galaxy are closely following the main sequence predicted in P. Popesso et al. (2023) and comparable to L. Clarke et al. (2024). However, on average they have lower metallicity at the given stellar mass compared to similar $4 < z < 10$ (F. D'Eugenio et al. 2024; K. E. Heintz et al. 2024; C. Marconcini et al. 2024; A. Sarkar et al. 2024; G. Venturi et al. 2024; X. Wang et al. 2024) galaxies (Figure 3(f)). Given the multiwavelength observations of MACS0416-Y1 and taking the estimate of molecular gas mass (estimated using [C II] dynamical decomposition) from G. C. Jones et al. (2024b), we measure the star formation efficiency (SFE) of the integrated galaxy as log SFR/$M_{H_2} = -8.49 \pm 0.24$ yr$^{-1}$, which is higher compared to the median SFE of local galaxies (D. Colombo et al. 2020) but comparable to the SFE of galaxies at cosmic noon (T.-M. Wang et al. 2022; Figure 3(h)). We note that the molecular gas mass estimated from L[C II] may be related to star formation and thus SFE (J. Kennicutt 1998). The ongoing star formation burst may lead to an overestimation of the molecular gas mass. Figure 3(g) further shows the position of MACS0416-Y1 in comparison to cosmic noon (T.-M. Wang et al. 2022) and local galaxies (L. Lin et al. 2019) on the molecular gas surface density–stellar mass surface density plane and the SFR surface density–molecular gas mass surface density plane. While MACS0416-Y1 is considerably more gas rich compared to the local and cosmic noon galaxies, it has comparable SFR surface density for a given gas mass surface density.

### 3.2. Physical Properties with Emission Line Ratios

With the NIRSpec prism, we observe MACS0416-Y1 using two three-shutter slits covering clumps 2 and 3 and part of clump 1 (Figure 2(a)). We use the emission line ratios to infer ISM properties. To accurately derive the physical properties of the galaxy using emission line ratios, we need first to correct the measured line fluxes for dust attenuation. We follow the Calzetti extinction law (D. Calzetti et al. 2000) and use the ratio of H$\beta$/H$\delta$ to correct the observed line fluxes. We measure H$\beta$/H$\delta$ = 2.12 compared to the Case B recombination value of 3.78 for $T_e = 1.6 \times 10^4$ K. We calculate nebular reddening $E(B-V)_{\rm gas} = 0.34$ and use the reddening-corrected line fluxes to calculate the following emission line ratios:

$$O3 = \log([\text{O\,III}]\lambda 4960, 5008/\text{H}\beta)$$
$$RO3 = \log([\text{O\,III}]\lambda 4960, 5008/[\text{O\,III}]\lambda 4363$$
$$R3 = \log([\text{O\,III}]\lambda 5008/\text{H}\beta)$$
$$O3Hg = \log([\text{O\,III}]\lambda 4363/\text{H}\gamma)$$
$$O33 = \log([\text{O\,III}]\lambda 5008/[\text{O\,III}]\lambda 4363)$$
$$R2 = \log([\text{O\,II}]\lambda 3727, 9/\text{H}\beta)$$
$$R23 = \log\left(\frac{[\text{O\,III}]\lambda 4960, 5008 + [\text{O\,II}]\lambda 3727, 9}{\text{H}\beta}\right)$$
$$O32 = \log([\text{O\,III}]\lambda 5008/[\text{O\,II}]\lambda 3727, 9)$$
$$Ne3O2 = \log([\text{Ne\,III}]\lambda 3869/[\text{O\,II}]\lambda 3727, 9)$$

We measure the electron temperature $T_e$([O III]) of the high-ionization O++ zone using emission line ratio [O III]$\lambda$4960, 5008/[O III]$\lambda$4363. In the NIRSpec prism data, [O III]$\lambda$4363 is blended with H$\gamma$. We calculate the expected H$\gamma$ flux based on the H$\beta$/H$\delta$ ratio and subtract the same from the combined [O III]$\lambda$4363+H$\gamma$ flux to estimate the [O III]$\lambda$4363 flux. Following D. C. Nicholls et al. (2020), we measure $T_e$([O III]) = 17634 ± 3900 K. This $T_e$ is comparable to $z \sim 6-8$ galaxies in, e.g., M. Curti et al. (2022), D. Schaerer et al. (2022), and R. L. Sanders et al. (2023). We also measure the hydrogen ionizing photon production efficiency of MACS0416-Y1 following the procedure described in A. Harshan et al. (2024) and find $\log \xi_{\rm ion} = 25.34 \pm 0.21$ ergs$^{-1}$ s$^{-1}$, which is comparable to $\xi_{\rm ion}$ found for galaxies at $z > 5$ (e.g., A. Harshan et al. 2024; S. Mascia et al. 2024; P. Rinaldi et al. 2024, and references within).

Next, we measure the electron density ($\ne$) of MACS0416-Y1 using the ratio [O III]$\lambda$88 $\mu$m/[O III]$\lambda$4960, 5008. We use the $L_{\rm [O\,III]\lambda 88\,\mu m} = 0.66 \pm 0.16$ Jy km s$^{-1}$ from Y. Tamura et al. (2019). Using PyNeb, assuming electron temperature $T_e = 18000 \pm 4000$ K as derived from line ratio [O III]$\lambda$4960, 5008/[O III]$\lambda$4363, we estimate $\ne = 100$ cm$^{-3}$, with an upper limit of <350 cm$^{-3}$ (Figure 3(d)).

Based on the electron density and temperature, we estimate the oxygen abundance using ratios O3 and R2. Following Y. I. Izotov et al. (2006), we estimate the ionic abundance as $12 + \log \text{O}^+/\text{H}^+ = 6.9 \pm 0.1$ and $12 + \log \text{O}^{2+}/\text{H}^+ = 7.7 \pm 0.1$. From the ionic abundances, we estimate the oxygen abundance as $12 + \log \text{O/H} = 7.8 \pm 0.2$, which is approximately 15% of the solar abundance, assuming a solar abundance of 8.6. Further, we calculate the ionization parameter using ratios O32 and Ne3O2. Following L. J. Kewley et al. (2019) and O32, we measure ionization parameter $\log U = -2.5 \pm 0.3$, and following E. M. Levesque & M. L. A. Richardson (2014) and using Ne3O2, we measure $\log U = -3.1 \pm 0.3$. We present the physical parameters estimated using emission lines in Table 2.

### 3.3. Emission Line Diagnostics

*O32-R23 Diagram.* The O32-R23 diagram is an important diagnostic diagram that relates the strength of the ionization field (through O32) and the gas-phase metallicity (through R23). Figure 3(a) shows the O32-R23 diagram of MACS0416-Y1 compared to literature values from the MOSDEF sample at $z \sim 2$ (R. L. Sanders et al. 2021), CEERS sample at $z > 5$ (R. L. Sanders et al. 2023; M. Tang et al. 2023), and JADES sample at $z > 5$ (A. J. Cameron et al. 2023). MACS0416-Y1 at $z = 8.317$ has higher O32 at its R23 compared to MOSDEF galaxies at $z \sim 2$ but has lower O32 compared to galaxies at R23 in the JADES sample at $z \sim 6$ and CEERS galaxies at $z \sim 7.7$. We find that MACS0416-Y1 has a weaker ionization field compared to galaxies at similar redshifts but higher than galaxies at $z > 5$ and $z \sim 2$. Meanwhile, it has higher metallicity compared to galaxies at similar redshifts but comparable to galaxies at $z > 5$ and $z \sim 2$. Thus we infer that MACS0416-Y1 is more evolved than its $z > 5$ counterparts.

*R2-R3 Diagram.* The R2-R3 diagram is also a useful diagnostic, especially in the absence of the typical diagnostic lines like [N II], [S II], and [O I]. Figure 3(b) shows the R2-R3 diagram of MACS0416-Y1 along with the comparison sample. A. J. Cameron et al. (2023) find in the JADES sample at $z > 6$ a median value of R2 of $-0.23 \pm 0.38$, which is lower than the value we find for MACS0416-Y1 (R2 = 0.19 ± 0.02). MACS0416-Y1 lies in the higher edge of R2 and R3 ratios compared to the high-redshift sample; however, MACS0416-Y1 has a lower R2 and higher R3 compared to the average sample from MOSDEF at $z = 2$. This again indicates that MACS0416-Y1 is a more evolved and metal-rich galaxy compared to the rest of the high-redshift comparison sample. We also note that MACS0416-Y1 is more comparable in R2-R3 to the sample of merging galaxies from G. Venturi et al. (2024) at $z > 6$.

We further overlay the R2-R3 diagram with shock models from M. G. Allen et al. (2008) in pink lines. The dashed line is at solar abundance, gas density ($n$) 1000 cm$^{-3}$, and shock velocity 500 km s$^{-1}$ and the solid line at velocity 1000 km s$^{-1}$. We find that the measured position of MACS0416-Y1 on the R2-R3 diagram can be described by a shock velocity of >1000 km s$^{-1}$. We also overlay the dusty narrow-line Seyfert models from B. A. Groves et al. (2004) in red at solar abundance. The red solid line and dashed lines show $\alpha = 1.7$ and $\alpha = 2.0$, respectively. We find that the R2 and R3 ratio of MACS0416-Y1 can be explained with dusty narrow-line Seyfert models.

### 3.4. Is MACS0416-Y1 a Possible Active Galactic Nucleus?

G. Mazzolari (2024) present a new active galactic nucleus (AGN) diagnostics using [O III]$\lambda$4363 auroral line. Similarly, R. Maiolino et al. (2023) also suggests a high O3Hg ratio as an indication of AGN. The [O III]$\lambda$4363 is sensitive to the electron temperature, and thus, high [O III]$\lambda$4363 flux and high electron temperature should indicate ionization sources other than the star formation. The position of MACS0416-Y1 on [O III]$\lambda$4363 diagnostics is shown in Figure 3(c). MACS0416-Y1 lies in the AGN-only region of O3Hg-Ne3O2, O3Hg-O32 diagnostic plots, and on the line delineating AGN to star-forming galaxies in the O3Hg-O33 plot (not shown). This indicates tentative evidence for the presence of AGN in MACS0416-Y1.

The presence of AGN is also indicated in the R2-R3 diagram (Figure 3(b)) by the dusty narrow-line Seyfert models from B. A. Groves et al. (2004). In our R100 prism spectrum, we do not find evidence for a broad-line region and find similar rest-frame velocity dispersions of 536 ± 10, 521 ± 10, and 515 ± 10 kms$^{-1}$ for H$\beta$, [O III]$\lambda$4961, and [O III]$\lambda$5007 lines, respectively. The small difference between the rest-frame velocities of H$\beta$ compared to [O III] lines is consistent with the change in dispersion per pixel for the NIRSpec prism. Further exploration of the gas kinematics for MACS0416-Y1 and the detection of a broad-line component using higher-resolution spectroscopy are required to make an inference in the existence of an AGN in MACS0416-Y1. This is beyond the scope of the NIRSpec prism data presented in this Letter.

### 4. Discussion and Summary

MACS0416-Y1, a $z = 8.317$ massive galaxy of stellar mass $1 \pm 0.07 \times 10^9 M_\odot$, has been a subject of various studies owing to its peculiar dust properties as seen from ALMA observations (T. J. L. C. Bakx et al. 2020; Y. Tamura et al. 2023; G. C. Jones et al. 2024b). Previous studies based on ALMA observations and HST imaging revealed that MACS0416-Y1 has a high dust temperature of >90 K and high molecular gas mass of $10^{10} M_\odot$. As part of the CANUCS survey, we observed MACS0416-Y1 with deep NIRCam imaging and NIRSpec spectroscopy. The resolved photometry from NIRCam data revealed three resolved clumps (C1, C2, and C3) along with an underlying galaxy halo (H) with magnification-corrected stellar masses $(M_\odot) = 3.9 \pm 0.1 \times 10^8$, $3.9 \pm 0.1 \times 10^7$, $1.5 \pm 0.3 \times 10^8$, and $2.5 \pm 2 \times 10^9$, respectively. With the SFR derived from JWST photometry and the molecular gas mass from ALMA, we measure the SFE of the galaxy and find that at $z = 8.317$, MACS0416-Y1 has comparable SFE to local and cosmic noon galaxies. We also measure the SFH and find simultaneous burst of star formation in the resolved clumps in MACS0416-Y1, indicating that MACS0416-Y1 is in a late-stage merger. Spatially resolved spectroscopy of MACS0416-Y1 is required to study the kinematics of the three resolved clumps to further strengthen the evidence for merger.

The emission line diagnostics of MACS0416-Y1 indicate that it is a metal-rich, highly evolved galaxy compared to the $z > 6$ sample, with metallicity similar to the average metallicity of the $z \sim 2$ sample. However, MACS0416-Y1 has a lower ionization parameter compared to the $z > 6$ sample but higher than the $z > 2$ sample. The closest sample of galaxies with comparable ionization properties and gas-phase metallicities is presented by G. Venturi et al. (2024), which is a sample of merging galaxies at $z > 6$ comparable to MACS0416-Y1 in stellar mass and SFR. The position of MACS0416-Y1 on the O3Hg-O32 plot indicates a possible presence of AGN

(G. Mazzolari 2024). With the prism spectrum, we find no significant difference between the velocity dispersion of [O III] and H$\beta$ emission lines, indicating a lack of high-velocity gas. However, further study of the broad-line region with higher-resolution spectroscopy is necessary to state the presence of AGN in MACS0416-Y1 conclusively.

A plausible physical scenario that explains the observations is a gas-rich major merger of two evolved stellar clumps inducing a star formation burst and possibly AGN activity. Such a star formation burst or AGN activity can explain the measured unusually higher dust temperature. This Letter presents an important step toward understanding the formation of massive evolved galaxies at the epoch of reionization (EoR). With a view of MACS0416-Y1 from UV to IR wavelengths, we present a view of galaxy assembly in the EoR. Further campaigns studying EoR galaxies in rest-frame IR are required to fully understand the evolution of dust and molecular gas properties. The consensus of dust and gas properties of a similar galaxy population, along with stellar properties, is required to fully understand the formation of first galaxies in the Universe.

### Acknowledgments

Some of the data presented in this Letter were obtained from the Mikulski Archive for Space Telescopes (MAST) at the Space Telescope Science Institute. The specific observations analyzed can be accessed via doi:10.17909/ph4n-6n76. This Letter also makes use of the following ALMA data: ADS/JAO. ALMA #2019.1.00343.S. ALMA is a partnership of ESO (representing its member states), NSF (USA), and NINS (Japan), together with NRC (Canada), NSTC and ASIAA (Taiwan), and KASI (Republic of Korea), in cooperation with the Republic of Chile. The Joint ALMA Observatory is operated by ESO, AUI/NRAO, and NAOJ.

A.H., M.B., G.R., R.T., V.M., and N.M. acknowledge support from the ERC Grant FIRSTLIGHT and Slovenian national research agency ARRS through grants N1-0238 and P1-0188. M.B. acknowledges support from the program HST-GO-16667, provided through a grant from the STScI under NASA contract NAS5-26555. This research was enabled by grant 18JWST-GTO1 from the Canadian Space Agency and funding from the Natural Sciences and Engineering Research Council of Canada. This research used the Canadian Advanced Network For Astronomy Research (CANFAR) operated in partnership by the Canadian Astronomy Data Centre and The Digital Research Alliance of Canada with support from the National Research Council of Canada the Canadian Space Agency, CANARIE, and the Canadian Foundation for Innovation. The Cosmic Dawn Center (DAWN) is funded by the Danish National Research Foundation under grant No. 140. We also thank the anonymous reviewer for their careful reading of the Letter and suggestions.

### Appendix A
### Parameters for SED Fitting

We list the parameters used for `DENSE BASIS` and `Bagpipes` SED fitting of resolved and integrated photometry in Table A1.

Table A1
Bagpipes and DENSE BASIS Model Parameters

| Parameter | Range | Prior |
|---|---|---|
| DENSE BASIS | | |
| $\log M^*/M_\odot$ | (5, 11) | flat |
| SFR | (−2, 5) | flat sSFR |
| Z | (−1, 2) | flat |
| z | (8.29, 8.33) | flat |
| $A_V$ | (0, 2) | exp |
| Bagpipes | | |
| dblplaw $\tau$ | (0, 15) | flat |
| dblplaw $\alpha$ | (0.01, 10,000) | log10 |
| dblplaw $\beta$ | (0.01, 1000) | log10 |
| $\log M^*/M_\odot$ | (6, 11) | flat |
| Z | (0.1, 2) | flat |
| $A_V$ | (0, 2) | flat |
| $\log(U)$ | −2 | N/A |

## Appendix B
## Cold Dust SED Modeling

We model the dust continuum with a MBB function given by

$$S^{\rm obs}_{\nu_{\rm obs}} = S^{\rm obs}_{\nu/(1+z)} = \frac{\Omega}{(1+z)^3}[B_\nu(T_{\rm dust}(z)) - B_\nu(T_{\rm CMB}(z))] \times (1 - e^{-\tau_\nu}), \quad (B1)$$

where $\Omega = (1+z)^4 A_{\rm gal} D_L^{-2}$ is the solid angle, and $A_{\rm gal}$ and $D_L$ are the surface area and luminosity distance of the galaxy, respectively (R. Tripodi et al. 2024). The dust optical depth is

$$\tau_\nu = \frac{M_{\rm dust}}{A_{\rm galaxy}} k_0 \left(\frac{\nu}{250 \text{ GHz}}\right)^\beta, \quad (B2)$$

with $\beta$ the emissivity index and $k_0 = 0.45$ cm$^2$ g$^{-1}$ the mass absorption coefficient (A. Beelen et al. 2006). The solid angle is estimated using the continuum mean size from Y. Tamura et al. (2023). The effect of the CMB on the dust temperature is given by

$$T_{\rm dust}(z) = ((T_{\rm dust})^{4+\beta} + T_0^{4+\beta}[(1+z)^{4+\beta} - 1])^{\frac{1}{4+\beta}}, \quad (B3)$$

with $T_0 = 2.73$ K. We also considered the contribution of the CMB emission given by $B_\nu(T_{\rm CMB}(z) = T_0(1+z))$ (E. da Cunha et al. 2013).

In our case, fixing the dust temperature, the model has only two fitting parameters: dust mass ($M_{\rm dust}$) and $\beta$. We explore the 2D parameter space using a Markov Chain Monte Carlo (MCMC) algorithm implemented in the EMCEE package (D. Foreman-Mackey et al. 2013). We assume uniform priors for the fitting parameters: $3 < \log(M_{\rm dust}/M_\odot) < 9$, $1.0 < \beta < 5.0$. The best-fit models have been obtained from MCMC with 20 chains, 2000 trials, and a burn-in phase of ∼100.

## ORCID iDs

Anishya Harshan https://orcid.org/0000-0001-9414-6382
Roberta Tripodi https://orcid.org/0000-0002-9909-3491
Nicholas S. Martis https://orcid.org/0000-0003-3243-9969
Gregor Rihtaršič https://orcid.org/0009-0009-4388-898X
Maruša Bradač https://orcid.org/0000-0001-5984-0395
Yoshihisa Asada https://orcid.org/0000-0003-3983-5438
Gabe Brammer https://orcid.org/0000-0003-2680-005X
Guillaume Desprez https://orcid.org/0000-0001-8325-1742
Jasleen Matharu https://orcid.org/0000-0002-7547-3385
Vladan Markov https://orcid.org/0000-0002-5694-6124
Lamiya Mowla https://orcid.org/0000-0002-8530-9765
Ghassan T. E. Sarrouh https://orcid.org/0000-0001-8830-2166
Marcin Sawicki https://orcid.org/0000-0002-7712-7857
Victoria Strait https://orcid.org/0000-0002-6338-7295
Chris Willott https://orcid.org/0000-0002-4201-7367

## References

Algera, H., Inami, H., De Looze, I., et al. 2024, MNRAS, 533, 3098
Algera, H. S. B., Hodge, J. A., Riechers, D., et al. 2021, ApJ, 912, 73
Allen, M. G., Groves, B. A., Dopita, M. A., Sutherland, R. S., & Kewley, L. J. 2008, ApJS, 178, 20
Asada, Y., Sawicki, M., Abraham, R., et al. 2024, MNRAS, 527, 11372
Asada, Y., Sawicki, M., Desprez, G., et al. 2023, MNRAS, 523, L40
Bakx, T. J. L. C., Tamura, Y., Hashimoto, T., et al. 2020, MNRAS, 493, 4294
Beelen, A., Cox, P., Benford, D. J., et al. 2006, ApJ, 642, 694
Brada, M., Strait, V., Mowla, L., et al. 2024, ApJL, 961, L21
Bradley, L., Sipőcz, B., Robitaille, T., et al. 2022, astropy/photutils: v1.6.0, Zenodo, doi:10.5281/zenodo.7419741
Brammer, G. 2022, msaexp: NIRSpec analyis tools, v0.3.4, Zenodo, doi:10.5281/zenodo.7313329
Brammer, G., & Matharu, J. 2021, gbrammer/grizli: Release 2021, v1.3.2, Zenodo, doi:10.5281/zenodo.5012699
Brammer, G. B., Marchesini, D., Labbé, I., et al. 2016, ApJS, 226, 6
Calzetti, D., Armus, L., Bohlin, R. C., et al. 2000, ApJ, 533, 682
Cameron, A. J., Saxena, A., Bunker, A. J., et al. 2023, A&A, 677, A115
Carnall, A. C., McLure, R. J., Dunlop, J. S., & Davé, R. 2018, MNRAS, 480, 4379
Carnall, A. C., McLure, R. J., Dunlop, J. S., et al. 2019, MNRAS, 490, 417
CASA Team, Bean, B., Bhatnagar, S., et al. 2022, PASP, 134, 114501
Casey, C. M., Akins, H. B., Shuntov, M., et al. 2024, ApJ, 965, 98
Chabrier, G. 2003, PASP, 115, 763
Clarke, L., Shapley, A. E., Sanders, R. L., et al. 2024, arXiv:2406.05178
Colombo, D., Sanchez, S. F., Bolatto, A. D., et al. 2020, A&A, 644, A97
Curti, M., Hayden-Pawson, C., Maiolino, R., et al. 2022, MNRAS, 512, 4136
Curtis-Lake, E., Carniani, S., Cameron, A., et al. 2023, NatAs, 7, 622
da Cunha, E., Groves, B., Walter, F., et al. 2013, ApJ, 766, 13
Desprez, G., Martis, N. S., Asada, Y., et al. 2024, MNRAS, 530, 2935
D'Eugenio, F., Cameron, A. J., Scholtz, J., et al. 2024, arXiv:2404.06531
Faisst, A. L., Fudamoto, Y., Oesch, P. A., et al. 2020, MNRAS, 498, 4192
Foreman-Mackey, D., Hogg, D. W., Lang, D., & Goodman, J. 2013, PASP, 125, 306
Groves, B. A., Dopita, M. A., & Sutherland, R. S. 2004, ApJS, 153, 75
Harshan, A., Bradač, M., Abraham, R., et al. 2024, MNRAS, 532, 1112
Heintz, K. E., Brammer, G. B., Watson, D., et al. 2024, arXiv:2404.02211
Infante, L., Zheng, W., Laporte, N., et al. 2015, ApJ, 815, 18
Iyer, K. G., Gawiser, E., Faber, S. M., et al. 2019, ApJ, 879, 116
Izotov, Y. I., Stasińska, G., Meynet, G., Guseva, N. G., & Thuan, T. X. 2006, A&A, 448, 955
Jones, G. C., Bunker, A. J., Telikova, K., et al. 2024a, arXiv:2405.12955
Jones, G. C., Witstok, J., Concas, A., & Laporte, N. 2024b, MNRAS, 529, L1
Jones, G. T., & Stanway, E. R. 2023, MNRAS, 525, 5720
Jullo, E., Kneib, J. P., Limousin, M., et al. 2007, NJPh, 9, 447
Kennicutt, J. 1998, ApJ, 498, 541
Kewley, L. J., Nicholls, D. C., Sutherland, R., et al. 2019, ApJ, 880, 16
Khusanova, Y., Bethermin, M., Le Fèvre, O., et al. 2021, A&A, 649, A152
Laporte, N., Streblyanska, A., Kim, S., et al. 2015, A&A, 575, A92
Levesque, E. M., & Richardson, M. L. A. 2014, ApJ, 780, 100
Lin, L., Pan, H.-A., Ellison, S. L., et al. 2019, ApJL, 884, L33
Lotz, J. M., Koekemoer, A., Coe, D., et al. 2017, ApJ, 837, 97
Ma, Z., Sun, B., Cheng, C., et al. 2024
Marconcini, C., D'Eugenio, F., Maiolino, R., et al. 2024, MNRAS, 533, 2488
Markov, V., Gallerani, S., Ferrara, A., et al. 2024, arXiv:2402.05996
Markov, V., Gallerani, S., Pallottini, A., et al. 2023, A&A, 679, A12
Mascia, S., Penterricci, L., Calabrò, A., et al. 2024, A&A, 685, A3
Mazzolari, G., Übler, H., Maiolino, R., et al. 2024, A&A, 691, A345
Mérida, R. M., Gómez-Guijarro, C., Pérez-González, P. G., et al. 2024, A&A, 686, A64

Mowla, L., Iyer, K., Asada, Y., et al. 2024, arXiv:2402.08696
Nicholls, D. C., Kewley, L. J., & Sutherland, R. S. 2020, PASP, 132, 033001
Parlanti, E., Carniani, S., Übler, H., et al. 2024, A&A, 684, A24
Peng, C. Y., Ho, L. C., Impey, C. D., & Rix, H.-W. 2010, AJ, 139, 2097
Popesso, P., Concas, A., Cresci, G., et al. 2023, MNRAS, 519, 1526
Rihtaršič, G., Bradač, M., Desprez, G., et al. 2024, arXiv:2406.10332
Rinaldi, P., Caputi, K. I., Iani, E., et al. 2024, ApJ, 969, 12
Robertson, B. E., Tacchella, S., Johnson, B. D., et al. 2023, NatAs, 7, 611
Sanders, R. L., Shapley, A. E., Topping, M. W., Reddy, N. A., & Brammer, G. B. 2023, ApJ, 955, 54
Sanders, R. L., Shapley, A. E., Jones, T., et al. 2021, ApJ, 914, 19
Sarkar, A., Chakraborty, P., Vogelsberger, M., et al. 2024, arXiv:2408.07974
Sarrouh, G. T. E., Muzzin, A., Iyer, K. G., et al. 2024, ApJL, 967, L17
Schaerer, D., Marques-Chaves, R., Barrufet, L., et al. 2022, A&A, 665, L4
Sommovigo, L., Ferrara, A., Carniani, S., et al. 2021, MNRAS, 503, 4878
Tamura, Y., C. Bakx, T. J. L., Inoue, A. K., et al. 2023, ApJ, 952, 9
Tamura, Y., Mawatari, K., Hashimoto, T., et al. 2019, ApJ, 874, 27
Tang, M., Stark, D. P., Chen, Z., et al. 2023, MNRAS, 526, 1657
Tripodi, R., Feruglio, C., Fiore, F., et al. 2024, A&A, 689, A220
Übler, H., Maiolino, R., Curtis-Lake, E., et al. 2023, A&A, 677, A145
Venturi, G., Carniani, S., Parlanti, E., et al. 2024, A&A, 691, A19
Viero, M. P., Sun, G., Chung, D. T., Moncelsi, L., & Condon, S. S. 2022, MNRAS, 516, L30
Wang, B., Fujimoto, S., Labbé, I., et al. 2023, ApJL, 957, L34
Wang, T.-M., Magnelli, B., Schinnerer, E., et al. 2022, A&A, 660, A142
Wang, X., Cheng, C., Ge, J., et al. 2024, ApJL, 967, L42
Willott, C. J., Doyon, R., Albert, L., et al. 2022, PASP, 134, 025002
Witstok, J., Jones, G. C., Maiolino, R., Smit, R., & Schneider, R. 2023, MNRAS, 523, 3119
Xiao, M., Oesch, P., Elbaz, D., et al. 2024, Natur, 635, 311